\documentclass[aps,prl,twocolumn,letterpaper,superscriptaddress,showpacs]{revtex4}
\usepackage{graphicx}
\usepackage{amsmath}
\usepackage{extarrows}

\begin{document}
\title{Create Dirac Cones in Your Favorite Materials}
\author{Chia-Hui Lin}
\altaffiliation{corresponding email: chia-hui.lin@stonybrook.edu}
\affiliation{Condensed Matter Physics and Materials Science Department,
Brookhaven National Laboratory, Upton, New York 11973, USA}
\affiliation{Department of Physics and Astronomy, Stony Brook University, Stony Brook, New York 11794, USA}
\author{Wei Ku}
\affiliation{Condensed Matter Physics and Materials Science Department,
Brookhaven National Laboratory, Upton, New York 11973, USA}
\affiliation{Department of Physics and Astronomy, Stony Brook University, Stony Brook, New York 11794, USA}

\date{\today}

\maketitle

\begin{figure}
    \centering
  \includegraphics[width=1\columnwidth,clip=true]{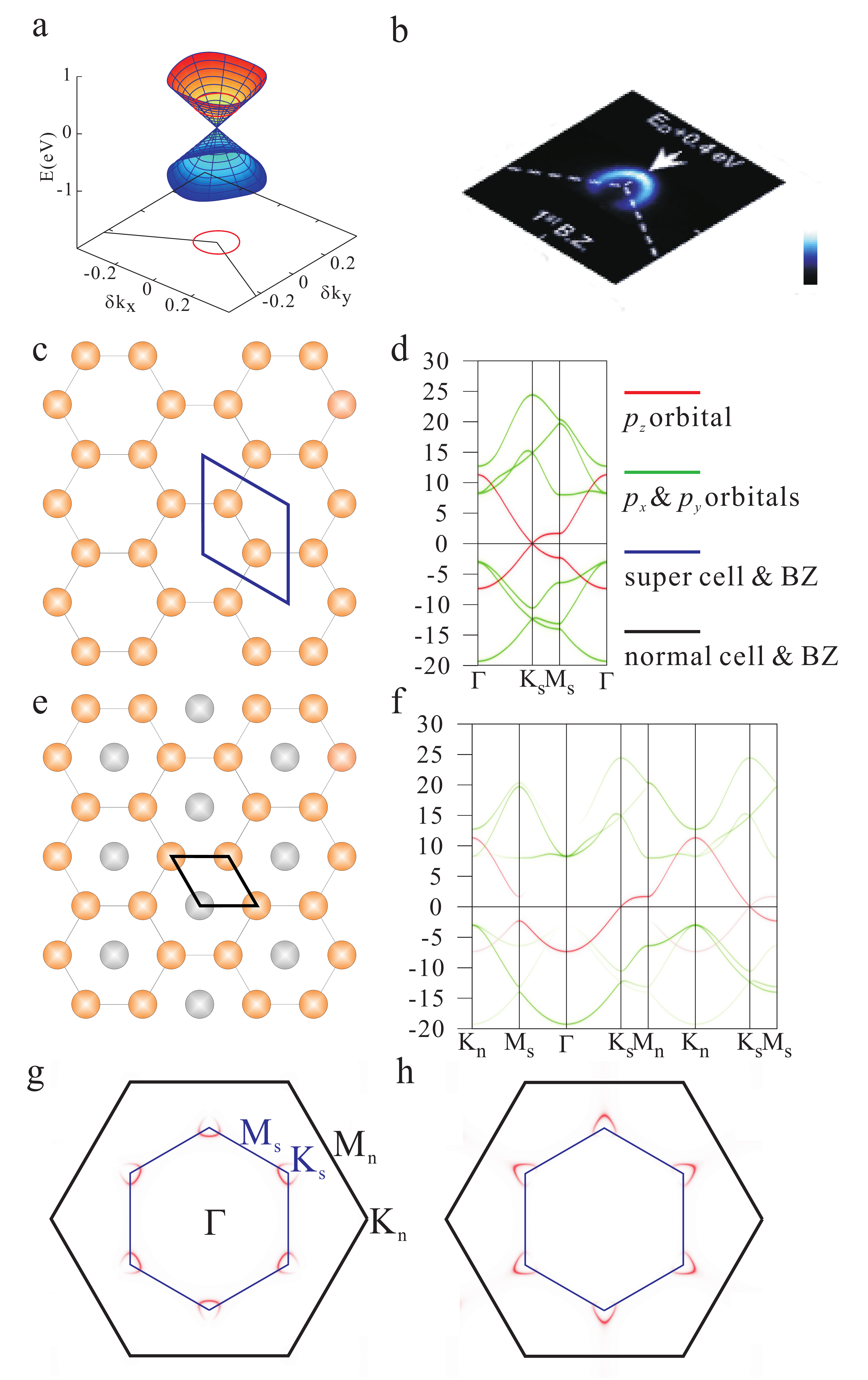}
  \caption{\textbf{New perspective of graphene lattice.} (a) Schematic diagram of linear dispersion of Dirac cones in graphene at the Brillouin zone corner (K point). On the base, the red circle represents the energy contour at 0.4 eV above the cone Fermi energy. (b) ARPES data from Ref. \cite{Hwang2011} at 0.4 eV above Dirac point. (c) The graphene honeycomb lattice of the blue unit cell (as the super cell). (d) The first-principle band structure with color code: red for $p_z$ and green for $p_x$ \& $p_y$ orbitals. The zero energy is chosen at the Fermi energy of graphene. The bands of two different color types are fully decoupled because the coupling is forbidden by the mirror symmetry with respect to the graphene sheet. (e) A reference triangular lattice of the black unit cell (as the normal cell) with the imaginary gray carbon, located at the hexagon center and treated as an vacancy. (f) The first-principle unfolded band structure in the normal cell basis. The thickness represent the magnitude of spectral weight of Green's function. The subscripts "s" and "n" are used to distinguish the high symmetry points defined for the super and normal cells. (g) and (h) represent the unfolded energy iso-surfaces at energy $= \pm$1 eV.
  }
  \label{fig:fig1}
\end{figure}

\textbf{
Realization of conically linear dispersion (termed Dirac cone as in Fig.~\ref{fig:fig1}(a)) has recently opened up exciting opportunities for high-performance devices that make use of the peculiar transport properties~\cite{Novoselov2004,Katsnelson2006,Novoselov2007,Du2008,Young2009,Gabor2011} of the massless carriers.
A good example of current fashion is the heavily studied graphene, a single atomic layered graphite.
It not only offers a prototype of Dirac physics in the field of condensed matter and materials science~\cite{Novoselov2005}, but also provides a playground of various exotic phenomena~\cite{Zhang2005,Bolotin2009,Du2009,Guinea2010}.
In the meantime, numerous routes have been attempted to search for the next "graphene"~\cite{Park2009,Asano2011,Kobayashi2007,Mori2010,Ran2009,Richard2010,Hasan2010}.
Despite these efforts, to date there is still no simple guideline to predict and engineer such massless particles in materials.
Here, we propose a theoretical recipe to create Dirac cones into anyone's favorite materials.
The method allows to tailor the properties, such as anisotropy and quantity, in any effective one-band two-dimensional lattice. We demonstrate the validity of our theory with two examples on the square lattice, an ``unlikely'' candidate hosting Dirac cones, and show that a graphene-like low-energy electronic structure can be realized. The proposed recipe can be applied in real materials via introduction of vacancy, substitution or intercalation, and also extended to photonic crystal~\cite{Huang2011}, molecular array \cite{Gomes2012}, and cold atoms systems~\cite{Tarruell2012}.
}

Let's first review an interesting observation of the Dirac cones~\cite{Wallace1947} in graphene via angle-resolved photoemission spectroscopy (ARPES)~\cite{Zhou2006,Bostwick2007, Hwang2011}, since the resolution of it would lead naturally to our proposed method.
Figure~\ref{fig:fig1}(b) shows an representative ARPES observation of the Dirac cone using out-of-plane polarized light.
Intriguingly, the observed cone appears incomplete even though a standard theory would indicate a complete Dirac cone, for example given by the intense red bands in Fig.~\ref{fig:fig1}(d).
This vanishing intensity is typically considered a ``matrix element effect'' of the measurement, and indicates a perfect destructive quantum mechanical interference~\cite{Shirley1995} between the two carbon atoms in the unit cell of the honeycomb lattice shown in Fig.~\ref{fig:fig1}(c).

An alternative and more straightforward perspective is to consider the honeycomb lattice as a triangular lattice with a periodic vacancy, as shown in Fig.~\ref{fig:fig1}(e).
Since there is only one (or sometimes zero) atom in this unit cell, the matrix element would just be the simple atomic form factor and the vanishing intensity is thus explicitly incorporated in the corresponding one-particle spectral function.
Indeed, Fig.~\ref{fig:fig1}(f-h) shows the resulting ``unfolded'' spectral function~\cite{Ku2010} with a ``incomplete'' Dirac cone having vanishing spectral weight in the lower part of the cone along the $\mathrm{\overline{K_s M_n}}$ path, and in the upper part along the $\mathrm{\overline{\Gamma K_s}}$ path, in perfect agreement with the experimental observation.

\begin{figure*}
     \centering
    \includegraphics[width=2\columnwidth,clip=true]{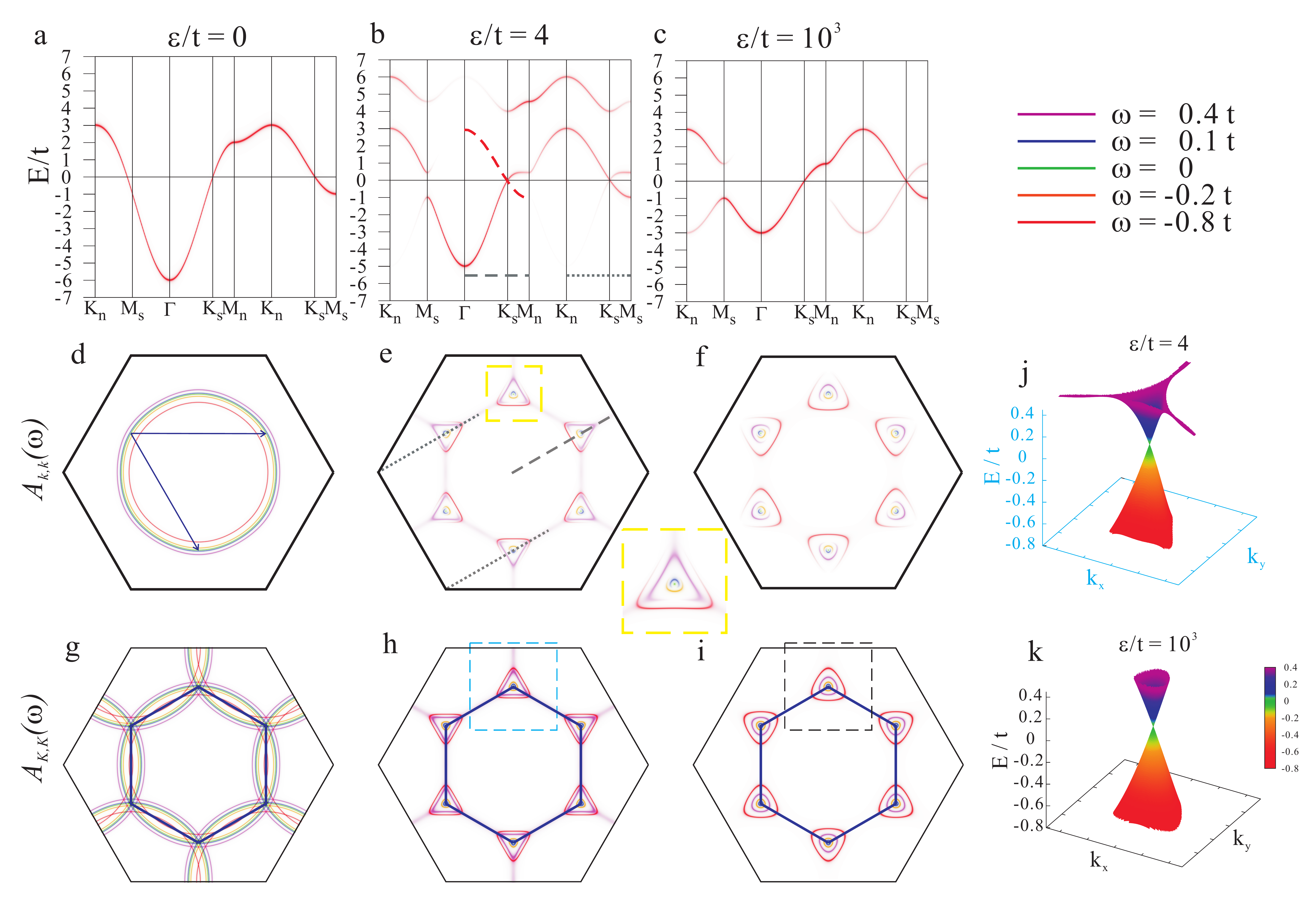}
  \caption{\textbf{Modeling of Dirac cone creation induced by CDW effects on the triangular lattice.} (a)-(c), unfolded band structure for $\varepsilon/t=$ 0, 4, $10^3$. (d) - (f) represent energy isosurfaces in the normal cell basis, namely $A_{k,k}(\omega)$, for three different $\varepsilon/t$. Different color stands for different energy cut with range from -0.8 t to 0.4 t. In order to show details of non-uniform spectral weight distribution, the yellow dashed square region in (e) is enlarged into an inset between the second and third row. (g) - (i) are the energy iso-surfaces plotted in super cell Brillouin zone. To directly visualize Dirac cones, (j) and (k) are the energy dispersion of the cyan and black dashed square in (h) and (i) respectively. The very flat band on top cone in (j) is responsible for the hexagonal shape of spectral function in (e) and (h). In (b) and (e), the two degenerate gray dotted $k$ path is coupled to the third gray dashed path by CDW potential. The red dashed curve in (b) refers to the missing folded spectral due to the special format of the potential.    }
  \label{fig:fig2}
\end{figure*}

This alternative ``one-carbon'' picture offers an interesting new way to understand the electronic structure of graphene, particularly the formation of the Dirac cones.
Fig.~\ref{fig:fig2} illustrates this with a reduced Hamiltonian, for clarity, that covers only the C-$p_z$ orbitals that define the relevant red band in Fig.~\ref{fig:fig1}.
Starting with a triangular lattice of carbon atoms with the same nearest inter-atomic coupling $t$ as in graphene, the corresponding band structure consists of a simple dispersion [Fig.~\ref{fig:fig2}(a)] and an almost circular Fermi surface [Fig.~\ref{fig:fig2}(d)] when the orbital is half-filled.
Upon raising the energy of the ordered vacancy sites by $\varepsilon$, the system is driven into a charge density wave (CDW) state, with most part of the Fermi surfaces (b)\&(c) gapped out, leaving six Dirac cones in the original one-carbon Brillouin zone (BZ) [denoted by solid black boundary lines in (d)(e)\&(f)].
As $\varepsilon$ grows, the effects is further enhanced and the cone becomes more and more symmetric [Fig.~\ref{fig:fig2}(j)(k)].
Finally, the perfectly symmetric cone of graphene is reproduced as the vacancy sites become forbidden, $\varepsilon\rightarrow\infty$.

Deeper physical and mathematical insights can be obtained by investigating the analytic structure of the above CDW potential in the momentum space:
\begin{equation}
V_{k,k+q} \propto \sum_{rr^{\prime}} V_{rr^{\prime}} e^{i k ( r^{\prime} -r)} e^{i q r^{\prime}},
\end{equation}
where $r$ and $r^\prime$ denotes the atomic sites, $k$ the crystal momentum and $q$ the CDW wave vector.
With only a local energy increase at the vacancy sites $V_{r,r^\prime}=\varepsilon \delta_{r,r^\prime}$, $V_{k,k+q}$ displays a very specific form within the subspace of the three states that couple together:
\begin{equation}
V_{k,k+q} =
\frac{\varepsilon}{3}
\left(
  \begin{array}{ccc}
  1 & 1 & 1 \\
  1 & 1 & 1  \\
  1 & 1 & 1
  \end{array}
\right)
\xlongrightarrow{diagonalization}
\left(
  \begin{array}{ccc}
  0 & 0 & 0 \\
  0 & 0 & 0  \\
  0 & 0 & \varepsilon
  \end{array}
\right).
\label{eqn:eqn2}
\end{equation}
Applying such a coupling to three states in Fig.~\ref{fig:fig2}(a)(d)\&(g) that happen to be degenerate, one finds a two-fold degeneracy is preserved in the resulting eigenvalues, forming the Dirac points, the tips of the Dirac cones.
Furthermore, with a slight deviation $\delta k$ from these $k$ points, the energy of the coupled states in (a) would differs by small amounts $a$ and $b$, both proportional to $\delta k$.
\begin{equation}
\frac{\varepsilon}{3}
\left(
  \begin{array}{ccc}
  1+\frac{3a}{\varepsilon} & 1 & 1 \\
  1 & 1+\frac{3b}{\varepsilon} & 1  \\
  1 & 1 & 1
  \end{array}
\right).
\end{equation}
Taking $|a| < |b| \ll \varepsilon$ for convenience, one finds linear dispersion developing around the Dirac point: $\frac{\varepsilon}{3}\{\frac{a}{2\varepsilon} +\frac{2b}{\varepsilon}, \frac{3a}{\varepsilon} -  \frac{3a}{8b} \frac{a}{\varepsilon} \}$.
Within such a one-carbon picture, we have demonstrated the formation of Dirac cones through an induced CDW of specifically structured CDW potential.
The above analysis also locates the Dirac cones in momentum space from the reference CDW-free system \textit{before} the CDW $q$ vectors to be applied are given.

The vanishing spectral intensity in Fig.~\ref{fig:fig1} is also easily understood with a similar analysis.
Along the $k$ path on which two of the three coupled states remain degenerate ($a=0$), for example two $k$-points on the $\overline{\mathrm{K_n K_s M_s}}$ path [gray dotted lines in Fig.~\ref{fig:fig2}(b)(e)] coupled to the third on the $\overline{\mathrm{\Gamma K_s M_n}}$ path [gray dashed line in Fig.~\ref{fig:fig2}(b)(e)], one of the resulting eigenvalues remains unchanged (zero): $\{3+\frac{b}{\varepsilon}, \frac{2b}{\varepsilon}, 0\}$, and the corresponding eigenvector consists of anti-bonding superposition of \textit{only} the two originally degenerate states.
Consequently, this band will not be folded to the third $k$ point (on the gray dashed line), where the corresponding spectral intensity must then vanish in Fig.\ref{fig:fig1}(g)\&(h).
The absence of folding intensity may be more clearly visualized in Fig.~\ref{fig:fig2}(a)-(c), where the band along the red path in (a) never appears in the green path, against naive expectation.
In essence, within this one-carbon picture, the vanishing spectral intensity is intimately tied to the formation of the Dirac cones.

The above generic description suggests a powerful and general recipe to create Dirac cones in any quasi-2D materials.
It would allow adding the exotic physics of Dirac particles and enhanced transport property into one's favorite materials that already hosts other useful characteristics. This is achieved by inducing a CDW state through introduction of impurities like vacancy, substitution, or intercalation.
These kinds of impurities generates impurity potential dominantly local at the impurity sites, thus giving a nearly constant CDW coupling in momentum space similar to Eq.~\ref{eqn:eqn2}.
The sample should then be synthesized/annealed to homogeneous distribution of the impurity, such that at higher enough concentration the impurities are expected to order reasonably well.
The key CDW wave vector would then be controlled by the impurity concentration, to couple $M$ $(M \geq 3)$ $k$-points in the clean system.
Since geometrically it is \textit{always} possible to couple at least three degenerate states with a regular dispersion, the above derivation would dictate the appearance of Dirac cones around the $M$ $k$-points of these degenerate states. In principles there can be multiple sets of such coupled $k$-points, in which case the number of Dirac cones would multiply. As long as the structure of the materials is stable against the not-so-small amount of impurity, this recipe offers a controllable and robust means to generate massless Dirac particles in almost any materials of interest.

\begin{figure*}
    \centering
  \includegraphics[width=2\columnwidth,clip=true]{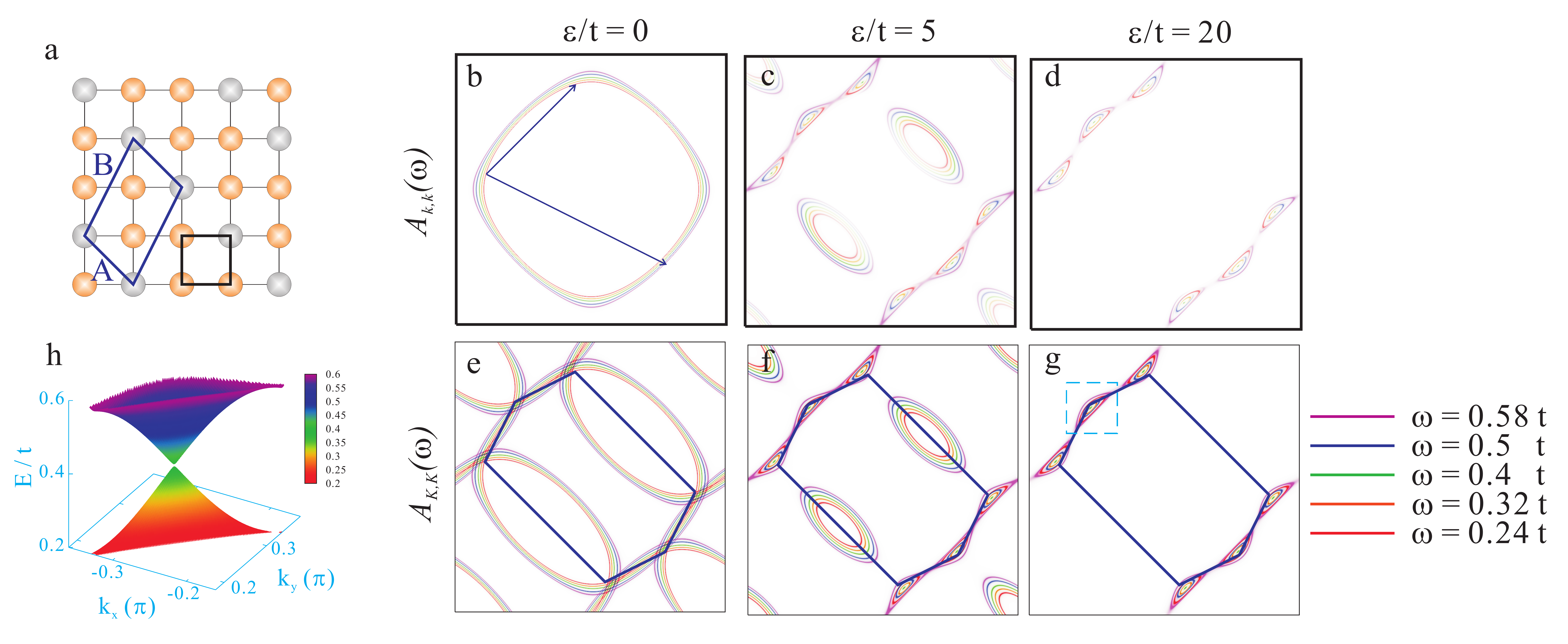}
  \caption{\textbf{Dirac cones on square lattice of 33\% ordered impurities (M=3)}. (a) The original square lattice of black normal cell and the blue super cell of $\{A=(1,\bar{1}),B=(1,2)\}$. The zero energy and next nearest hopping are chosen as the on-site energy of the unperturbed orange atoms and 0.4 $t$ respectively. The on-site energy of the gray atom is lifted to $\varepsilon$. (b) - (d) are the unfolded energy isosurfaces plotted in normal cell basis for $\varepsilon/t=$0, 5, 20. Different color represents different energy cut with range from 0.24 t to 0.58 t. (e) - (g) are energy isosurface in super cell basis. (h) is the energy dispersion of the region of cyan dashed square in (g). }
  \label{fig:fig3}
\end{figure*}

To demonstrate the general validity of this recipe, let us take a 2D one-band system with a common square lattice and introduce 33.3\% substitution of the atoms uniformly across the system.
Figure~\ref{fig:fig3} demonstrates creation of six anisotropic Dirac cones in the original BZ (two in the new reduced BZ).
In the case with weaker impurity potential $\varepsilon$ (Fig.~\ref{fig:fig3}(c)\&(f)), three regular ''electron pockets'' remain at the chemical potential.
In general, contribution to transport properties from the normal massive carriers on these pockets should be overwhelmed by those of the massless Dirac carriers, and thus does not cause any serious concern.
These normal carriers can often be removed by gapping out the pockets with a stronger impurity potential (for example with vacancy), as shown in Fig.~\ref{fig:fig3}(d)\&(g).
Obviously, the closer the reduced BZ is to the original Fermi surface, the more effectively the impurity can gap out these Fermi pockets.

\begin{figure*}
    \centering
  \includegraphics[width=2\columnwidth,clip=true]{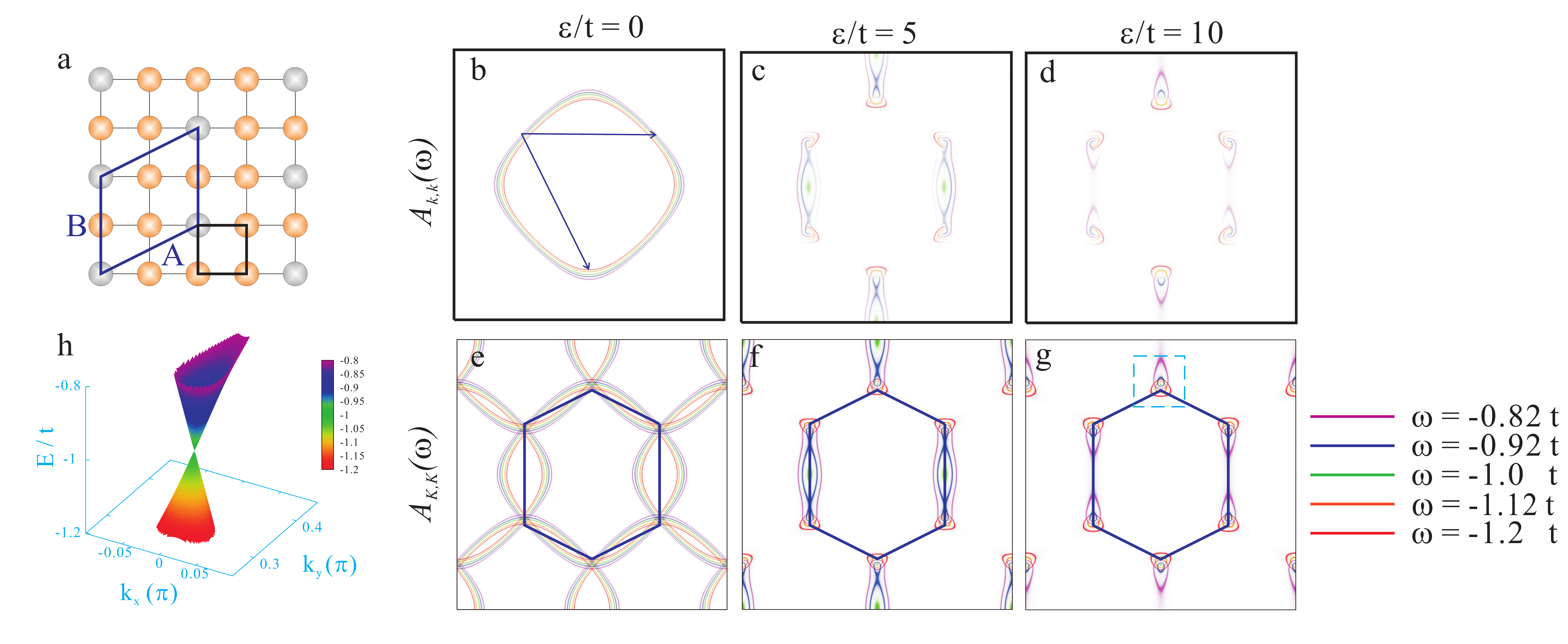}
  \caption{\textbf{Dirac cones on square lattice of 25\% ordered impurities (M=4)}. (a) The supercell of $\{A=(2,1),B=(0,2)\}$. The graph convention follows Fig.~ref{fig:fig3}. (b) - (d) are the energy isosurfaces plotted in normal cell basis for $\varepsilon/t=$0, 5, 20. Different color represents different energy cut with range from -1.2 t to -0.82 t. (e) - (g) are energy isosurfaces in super cell basis. (h) is the energy dispersion of the region of cyan dashed square in (g). }
  \label{fig:fig4}
\end{figure*}

One very interesting utilization of our recipe is to try to create a graphene-like electronic structure from a square lattice, rather than hexagonal structure.
Figure~\ref{fig:fig4} shows that introducing 25\% uniformly distributed impurities can indeed create six Dirac cones in the original BZ that resemble very much those in graphene.
Out of four coupled $k$ points, only three of them are degenerate (M=3).
The state at the fourth $k$-point has a much higher energy and thus does not affect the other three in any significant manner.

Finally, we clarify a few common practical aspects in real materials. First, it is obviously impossible to have a perfect order in the impurity. Luckily, the linearly vanishing density of states to the Dirac point and its massless feature would render the disorder effects of weak imperfection rather insignificant.
In essence, Dirac carriers are quite immune to disorder effects in general~\cite{Neto2009}.
Second, our analysis above takes a simple limit that the impurity potential is dominantly local in the impurity site.
While this is indeed the case for almost all the impurities (and certainly can be controlled by choosing the atoms for substitution), a small non-local influence of the impurity is expected in real materials.
Such a small non-local effect would introduce a small $k$-dependence of the coupling, and thus a small variation of the off-diagonal elements of Eq.~\ref{eqn:eqn2}.
This in turn would introduce a second order correction $\delta$ to the resulting eigen-energy and possibly lift the degeneracy of the Dirac point with a small gap. At the temperature/energy scale larger than $\delta$ one would still observe Dirac linear dispersion.
On the other hand, this offers an additional engineering degree of freedom to create a controlled semi-conducting gap that itself can have great deal of applications~\cite{Neto2009}. 

Third, for system with more 3D like dispersion, our mathematical argument would produce Dirac cone structure in the 2D CDW plane, but with Dirac point connected in the direction perpendicular to the plane.
This leads to an interesting situation of carriers having zero mass in the plane, but an almost infinite mass in the out-of-plane direction, essentially promoting a 2D characteristic.
Finally, the created Dirac cone might not always occurs at the chemical potential in real material. This can be overcome by the standard technique of doping or applying gate voltage.

\section{Acknowledgements}
Work funded by the U S Department of Energy, Office of Basic Energy Sciences DE-AC02-98CH10886 and by DOE-CMCSN.

\begin{widetext}
{\Large\bf Supplementary information}

\vspace{10mm}

In the supplementary information, we will provide computational details about the first-principle results graphene and band structure unfolding mentioned in the main text. Standard density functional theory (DFT) calculation is performed by using full potential linearized augmented plane wave method with local density approximation implemented in the WIEN2k package \cite{sWien2k} version 10.1. To simulate the quasi-two-dimensional properties of graphene of honeycomb lattice, the unit cell, containing two atoms, is chosen with 1.42 $\mathrm{\AA}$ inter-carbon distance and 10 $\mathrm{\AA}$ inter-layer distance vertically. This forms a crystal structure of P6/mmm space group. All input settings follow the default values as $R_{mt}K_{max} = 7$ and $l_{max} = 10$. The $k$ point mesh of $21\times 21 \times 4$ is used to reach convergent ground state density. Then the symmetry-respecting Wannier functions of carbon $s$ and $p$ characters would be constructed within [-20:30] eV low energy Hilbert space based on first-principles dispersion \cite{sKu2002}.

As mentioned in the main text, it is physically meaningful to represent the electronic band structure in the reference system corresponding to the one-carbon unit cell on triangular lattice. Indeed, this unfolding technique helps us to reveal the important symmetry breaking effect: the vacancy-induced charge density wave leads to the formation of Dirac cones. The basic idea of our unfolding method \cite{sKu2010} is to simply represent the energy ($\omega$) dependent one-particle spectral function, imaginary part of Green's function, of the real system (honeycomb graphene) by using the basis from reference system (triangular carbon lattice): $A_{kn,kn}(\omega) = \sum_{KJ}|\langle kn| KJ\rangle|^2 A_{KJ,KJ}(\omega)$,
where $K$/$k$ denotes the crystal momentum of the graphene/triangular lattice, $J$ the band index, and $n$ the Wannier orbital index (like $s$ and $p$ orbitals). This change of basis is made simple with the use of the above symmetry-respecting Wannier functions \cite{sKu2002}. For the demonstration purpose in the main text, we show both $A_{KJ,KJ}(\omega)$ and $A_{kn,kn}(\omega)$ to illustrate the clearer morphology and spectral weight distribution of the Dirac cone respectively.

\end{widetext}

\end{document}